\documentclass{PoS}

\let\oldbibliography\thebibliography
\renewcommand{\thebibliography}[1]{\oldbibliography{#1}
\setlength{\itemsep}{0pt}}

\title{Novel Signals from Neutron Star Mergers at 511 keV}

\ShortTitle{Novel Tracers of Neutron Star Mergers}

\author{\speaker{Volodymyr Takhistov} \\
        Department of Physics and Astronomy, University of California, Los Angeles\\
 Los Angeles, CA 90095-1547, USA\\
        E-mail: \email{vtakhist@physics.ucla.edu}}

\abstract{Synergetic observations of multi-band coincidence signals from merging neutron stars have definitively marked the significance of multi-messenger astronomy. We present a new generic signature of neutron star mergers, positron emission and the associated 511 keV radiation, produced from ejected neutron-rich radioactive merger material. Accounting for historical neutron star mergers within the Milky Way allows to readily explain the origin of the long-observed 511 keV emission line from the Galactic Center. Further, we draw a direct link between heavy element  production ($r$-process nucleosynthesis) and 511 keV emission, which signifies the surprising recent observations of Reticulum II ultra-faint dwarf spheroidal galaxy as a smoking gun of our proposal. This novel tracer of neutron star mergers provides a distinct handle for exploring binary merger history.}

\FullConference{36th International Cosmic Ray Conference -ICRC2019-\\
		July 24th - August 1st, 2019\\
		Madison, WI, U.S.A.}

\begin{document}

\section{Introduction}

Recent synergetic observations of coincident gravitational as well as electromagnetic signals from a binary neutron star (NS-NS) merger have strongly reaffirmed the significance of multi-messenger astronomy~\cite{GBM:2017lvd}.~Dense neutron-rich material ejected during the merger provides a favorable stage for $r$-process nucleosynthesis~\cite{Kasen:2017sxr,Pian:2017gtc,Drout:2017ijr} and powers electromagnetic ``kilonova'' transients~\cite{Li:1998bw,Metzger:2016pju}. $R$-process is a key production mechanism for heavy elements, such as gold and uranium, in astrophysics~\cite{Horowitz:2018ndv}. Here, copious amounts of neutrons rapidly capture on a seed nuclei before they can decay, allowing for a build-up of a heavy element with a large atomic number. Decompression and nuclear heating of the expanding and decaying radioactive ejecta material results in associated electromagnetic kilonova afterglow.~Similar type of emission is also expected of a neutron star-black hole (NS-BH) merger~\cite{Kawaguchi:2016ana}. It is clear that these events have definitely occurred within the Milky Way throughout the cosmological history.

For several decades now, a strong and sustained 511 keV emission line signal has been observed within the Galactic Center (GC)~\cite{Johnson:1972,Leventhal:1978}, with detailed measurements performed by the SPI spectrometer aboard the INTEGRAL satellite~\cite{Knodlseder:2005yq,Siegert:2016ijv}.
While a significant signal flux has been reported from the bulge component of the Galaxy, the disk component also displays non-negligible activity~\cite{Siegert:2015knp}. 
The observed 511 keV signal is consistent with photon emission from electron-positron annihilations through formation of intermediate positronium bound states, occurring at a rate
$\Gamma_p (e^+e^-\rightarrow \gamma\gamma) \sim 
 10^{50}\, {\rm yr}^{-1}$. As the positrons can readily annihilate in-flight directly, requiring that they cool and form positronium restricts their energies to lie below $\sim 3$ MeV~\cite{Beacom:2005qv}. The origin of these positrons remains an open question and many proposals have been put forth~\cite{Prantzos:2010wi}, including standard astrophysical sources (e.g.~pulsars winds~\cite{Wang:2005cqa}, nucleosynthesis from supernovae and massive stars~\cite{Prantzos:2010wi, Perets:2014mba,Alexis:2014rba,Milne:2001zs}, gamma-ray bursts~\cite{Bertone:2004ek}) as well as those based on beyond the Standard Model physics (e.g. primordial black holes\footnote{The 511-keV signal has been also suggested as a potential source of constraints on radiating primordial black holes~(e.g.~\cite{DeRocco:2019fjq,Laha:2019ssq}).}  disrupting compact stars~\cite{Fuller:2017uyd}, WIMP particle dark matter annihilations/de-excitations~\cite{Boehm:2003bt,Finkbeiner:2007kk,Pearce:2015zca}). In addition to Galactic Center, 511-keV emission has been also recently reported from ultra-faint spheroidal galaxies~\cite{Siegert:2016ijv}. Interestingly, the observed emission is particularly strong from Reticulum II, which also displays a large abundance of heavy elements, signifying $r$-process nucleosynthesis~\cite{Ji:2015wzg}.

 We will demonstrate, following Ref.~\cite{Fuller:2018ttb}, how NS-NS and NS-BH mergers are expected to generically produce copious amounts of thermal positrons, which originate from expanding radioactive merger ejecta material. The associated 511-keV signal from past mergers can readily explain the observed 511-keV excess in the Galactic Center. Our proposal provides a natural link between $r$-process nucleosynthesis and 511 keV radiation, which signifies the recent observations of Reticulum II as a smoking gun.
 Furthermore, we suggest that 511-keV radiation can be employed in future studies as a novel tracer of historical mergers.
  
 \section{Merger positron emission}
 
\begin{figure*}[ht!]
	\resizebox{\textwidth}{!}{
		\includegraphics{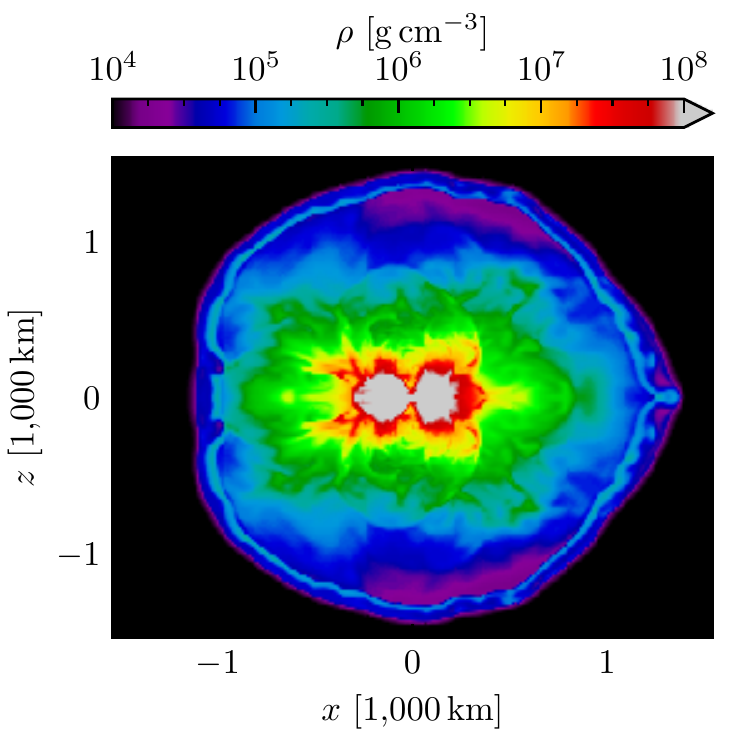}
    	\includegraphics{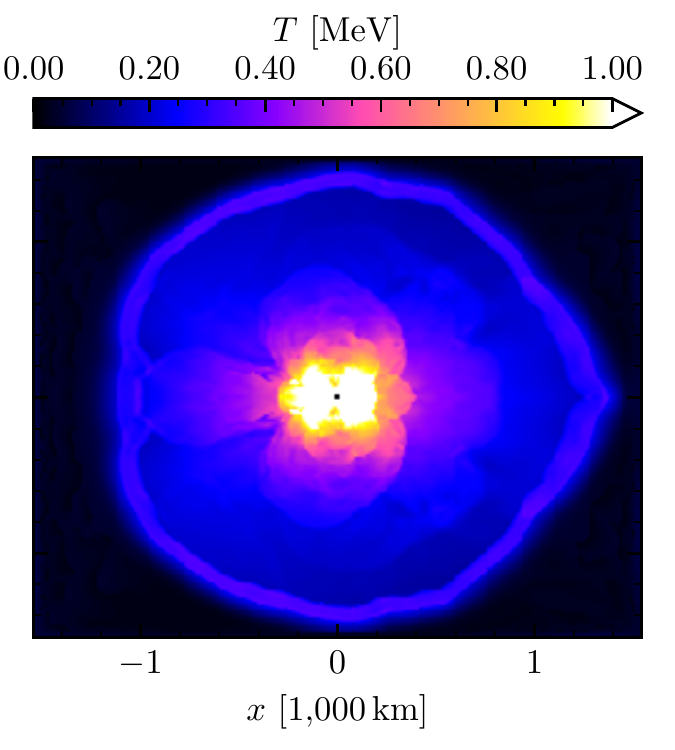}
    	\includegraphics{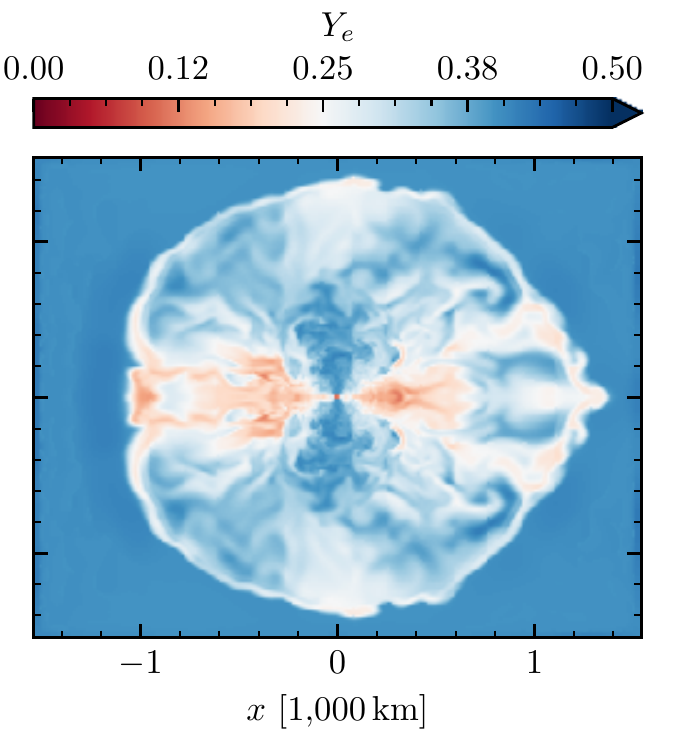}
    }
	\caption{Density $\rho$ (left), temperature $T$ (center) and electron fraction $Y_e$ (right) profiles of the ejected material at $t = 10$~ms after a typical neutron star merger. Simulation results reproduced from Ref.~\cite{Fuller:2018ttb}.}
	\label{fig:ejecta} 
\end{figure*}

Early stages of ejecta evolution from binary neutron stars mergers can be readily tracked with numerical relativity. On Fig.~\ref{fig:ejecta} we display some of the key quantities characterizing the ejecta  10~ms after the merger of a typical binary system: temperature $T$, density $\rho$ and electron fraction $Y_e$. Since the ejecta temperature is $\mathrm{O}(1)$ MeV, copious production of non-relativistic thermal positrons will take place. The associated positron number density $n_p (T)$ can be estimated from the Boltzmann distribution as
\begin{equation} \label{eq:boltz}
n_p (T) = 2 \Big(\frac{m_e T}{2 \pi}\Big)^{3/2}e^{-m_e/T}~,
\end{equation}
where $m_e = 0.511$ MeV is the positron mass.
Neutron stars are typically highly magnetized and magnetic fields are rampant throughout the expanding ejecta as well. However, ejecta magnetic confinement cannot be perfect, especially since the magnetic fields are randomized. Hence, some of the produced positrons are bound to leak out. While a full detailed analysis of the system would be highly nontrivial, similar studies for ejecta emission from supernovae suggest that up to $\mathrm{O}(10)\%$ of all positrons can escape.
 
\begin{figure*}[ht!]
 \centering
	\resizebox{0.5\textwidth}{!}{
		\includegraphics{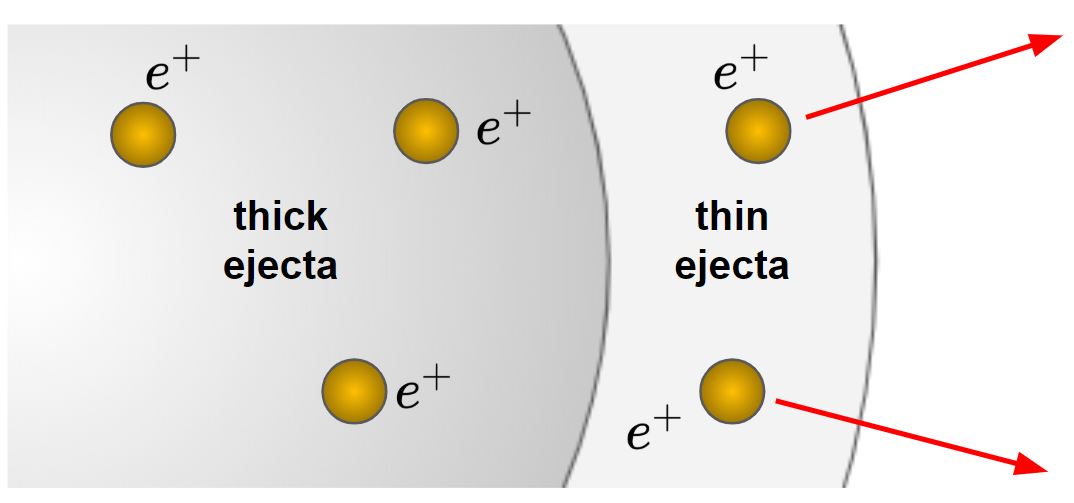}
    }
	\caption{Emission of positrons from the ``optically thin'' outer layers of expanding merger ejecta.}
	\label{fig:poseject} 
\end{figure*}

For positrons to escape, the outer layers of expanding ejecta must be ``optically thin'' (i.e. optical depth $\tau_e \lesssim 1$). Based on general physical arguments (e.g. \cite{Qian:1993dg}), the boundary between the outskirts of the ejecta gas and vacuum is a smooth transition through rarefication of layers. Hence, collection of the outermost ejecta layers can be approximated by a thin ``atmospheric layer'' that is exponentially decreasing in density. This layer lies below the resolution available in merger simulations which naively indicate that all of the resulting ejecta layers appear to be dense and initially optically thick to positrons (see Fig.~\ref{fig:ejecta}). However, when the atmospheric layer is accounted for, due to decreasing density, at any given moment in time there will exist a collection of outermost layers in the optically thin regime. This allows for some positrons to escape, as depicted on Fig.~\ref{fig:poseject}.

Due to a network of complex nuclear reactions~\cite{Rosswog:2013kqa}, the ejecta remains heated to temperatures of $\mathrm{O}(0.1)$ MeV for a duration of about $t_e \sim 1$~s after the merger. Subsequent cooling of the expanding ejecta leads to a dramatic decrease in the amount of associated positrons $n_p (T)$. Throughout $t_e$, the emission is dominated by the outermost layers and the total amount of emitted positrons can be approximated as
\begin{equation}
    N_p = n_p (T) S v_e t_e \simeq 5 \times 10^{58}~,
\end{equation}
where $S \simeq 4 \pi (v_e t_e)^2$ is the emitting surface area and $v_e \sim 0.8$\footnote{Units of $c = 1$ are assumed throughout.}~denotes the positron velocity.

\section{511 keV radiation}

Produced positrons will annihilate, resulting in emission of 511 keV radiation. Assuming neutron star binary merger rate of $R_{\rm MW} \simeq (10^{-2} - 10^2)$ Myr$^{-1}$ in the Milky Way, consistent with LIGO observations~\cite{Abbott:2016ymx,Mapelli:2018wys,Chruslinska:2017odi}, the average positron emission rate is approximately
\begin{equation}
    \Gamma_p = N_p R_{\rm MW} \simeq 5 \times 10^{50-54}~{\rm yr}^{-1}~.
\end{equation}
Hence, emission from neutron star mergers can readily address the observed 511 keV GC excess. We note that there is a sizable uncertainty in our estimates, which could be affected by a variety of factors (e.g. magnetic field geometry).

A general picture of the 511 keV signal morphology associated with mergers can be understood from considerations of diffusion, related time-scales as well as source distribution. 
Positrons of $\sim$MeV energies generated from the ejecta will diffuse within the interstellar medium to distances of $r_d \simeq \mathrm{O}(0.1)$ kpc~\cite{Jean:2009zj} over diffusion time of $t_d \simeq~ 10^{7-8}$ yrs~\cite{Bertone:2004ek}. Since the time-scales associated with mergers $t_m = 1/R_{\rm MW} \simeq 10^{4-5}$ yrs are significantly lower than $t_d$, past mergers have sufficient time to populate the $\mathrm{O}(1.5)$ kpc area within the Galactic bulge seen to shine in 511 keV. Observations of non-negligible 511 keV signal component from the Galactic disk suggest that the signal origin is related to star-formation and favors binary mergers over some of the alternative explanations, such as dark matter. While at first glance 511 keV emission from supernovae would also seem promising, it is difficult to reconcile disk emission with the expected supernova distribution~\cite{Prantzos:2010wi}.
In the case of neutron stars, however, binary kicks~\cite{Berger:2013jza} imply that a sizable disk component could be expected. On the other hand, since positron propagation is highly sensitive to magnetic fields and gas density~\cite{Bertone:2004ek}, we do not envision a significant signal component coming from the halo.

Recent observations of ultra-faint dwarf spheroidals indicate that Reticulum II shows a particularly strong emission in 511 keV~\cite{Siegert:2016ijv}. The same system also displays a significant abundance of heavy elements, typically associated with $r$-process nucleosynthesis, and it has been argued that a rare historical event is responsible~\cite{Ji:2015wzg}. These surprising observations constitute an obvious smoking gun of our proposal, which naturally links $r$-process nucleosynthesis with subsequent emission of 511 keV radiation originating from a rare event.

Emission of 511 keV radiation could be also employed for indirect detection of neutron star mergers. While supernova remnants are also expected to emit in 511 keV, the amount of ejected radioactive material associated with a supernova explosion is typically far lower than from a merger.~Hence, 511 keV emission associated with a supernova is expected to be significantly weaker. An analyses of 511 keV ``hot-spots'' could thus aid in distinguishing between supernova remnants and historic neutron star mergers.
 
\section{Summary}
\label{sec:summ}

A typical neutron star merger event results in a complex multi-messenger signal. We have shown that positron emission from expanding radioactive ejecta as well as the associated 511 keV radiation are generically expected from NS-NS and NS-BH mergers. Historic neutron star mergers allow for a natural explanation of the origin of the long-standing 511 keV emission signal observed from the Galactic Center.
Joint observations of significant heavy element abundance as well as 511 keV emission from Reticulum II ultra-faint dwarf spheroidal galaxy provide a smoking gun signature of our proposal, which naturally links $r$-process nucleosynthesis and the subsequent 511-keV emission originating from a rare event. This novel multi-messenger signal component could be used as a tracer of neutron star merger history.
 
\subsubsection*{Acknowledgments}
We thank the organizers of ICRC-2019 for the opportunity to present our results. The work of V.T. was supported by the U.S. Department of Energy (DOE) Grant No. DE-SC0009937.

\bibliography{511ns}
\addcontentsline{toc}{section}{Bibliography}
\bibliographystyle{JHEP}

\end{document}